\documentclass[11pt,twocolumn]{article}
\usepackage[utf8]{inputenc}
\usepackage{lipsum}
\usepackage[a4paper,margin=1.8cm,top=2cm,bottom=2.5cm]{geometry}

\usepackage[affil-it]{authblk}
\usepackage{blindtext}
\usepackage{hyperref}
\usepackage{url}
\usepackage{authblk}

\newcommand*{\doi}[1]{\href{https://doi.org/\detokenize{#1}}{doi: \detokenize{#1}}}

\usepackage{tabularx}
\usepackage{graphicx}
\usepackage{float}
\floatstyle{plaintop}
\restylefloat{table}

\usepackage{multicol}

\usepackage{fancyhdr}

\fancypagestyle{firstpage}{%
  \lfoot{\small \textit{Submitted for the requirements of MPhys at University of Oxford}}
  \rfoot{\small \textit{March 2, 2021}}
  \cfoot{}
  \fancyhead{}{}
}

\usepackage{enumitem}
\usepackage{gensymb}

\usepackage{amsmath}
\usepackage{array}
\usepackage[sort, numbers]{natbib}

\usepackage{caption}
\usepackage{changepage}

\makeatletter
\newbox\abstract@box
\renewenvironment{abstract}
  {\global\setbox\abstract@box=\vbox\bgroup
     \hsize=\textwidth\linewidth=\textwidth
    \normalsize
    \hrule%
    \vspace{0.5cm}
    {\hspace{0.7em} \bfseries \abstractname\vspace{-.5em}\vspace{\z@}}%
    \quotation}
  {\endquotation\egroup}
\expandafter\def\expandafter\@maketitle\expandafter{\@maketitle
  \ifvoid\abstract@box\else\unvbox\abstract@box\if@twocolumn\vskip1.5em\fi\fi}
\makeatother

\begin{document}

\newpage
\twocolumn
\pagenumbering{arabic}
\title{\textbf{Beam delivery systems for linac-based \\ proton therapy}}
\author[1]{Titus S. Dascalu\thanks{E-mail: t.dascalu19@imperial.ac.uk}}
\author[2,3]{Suzanne L. Sheehy}

\affil[1]{Department of Physics, Imperial College London, London, SW7 2AZ, UK}
\affil[2]{Department of Physics, University of Oxford, Oxford, OX1 3RH, UK}
\affil[3]{School of Physics, University of Melbourne, Melbourne, Victoria 3010, Australia}
\date{}

\begin{abstract}
\vspace{0.2cm}
\normalsize
\noindent
This report presents a design of a gantry for proton therapy based on the concept of adiabatic transition. The use of fixed-field alternating gradient magnets allows a large momentum acceptance and supports fast energy modulation. The optical performance of the gantry has been analysed using a beam tracking code. Several optimisations of the lattice and transition sections have been investigated to reduce size and ensure applicability to pencil beam scanning. Matching of the full energy range results in an increase in the size of the gantry, but reduces the weight and cost significantly compared to those that pertain to facilities in operation.  

\vspace{0.4cm}
\vspace{0.4cm} \\
\hrule
\vspace{0.3cm}
\end{abstract}

\maketitle
    \thispagestyle{firstpage}

\section{Introduction}

From the initial proposition of R. Wilson \cite{Wilson} in 1946 regarding the use of accelerated protons for radiotherapy, ion beams have proven to be an effective tool for cancer treatment. The key advantage of protons over X-rays is the precision with which the radiation dose can be delivered to the target tissue~\cite{PAGANETTI201377}. In the context of a rising world population and an increase in the complexity of the disease, there is a prompt need for novel techniques to be developed towards treating patients on a large scale~\cite{Atun2015} and with the high precision necessary for a better quality of life. By the end of 2019, around 260,000 patients \cite{ptcog} have been treated with particle radiotherapy at about 120 therapy facilities in operation worldwide. Approximately 85\% of these patients were treated with proton beams, while for the others, biologically more effective~\cite{Karger_2017} light-ion beams were used.

Hadrons are preferred since they have a significant increase in the dose at the end of their range \cite{CYRSP245}. These individual `Bragg peaks' can be delivered sequentially to obtain a spread out region of uniform energy deposition. While passive methods based on beam scattering are still in use, the state of the art technique is the so-called pencil beam scanning \cite{Giordanengo:2018tqo} -- the beam is swept across the transverse plane delivering the dose in a grid of spots. This is considered the optimal method since it removes the need for patient-specific devices and minimises the radiation of healthy tissues. Also, the parallel advance of imaging techniques makes it suitable for future treatment of moving organs. 

The main challenge in providing this treatment to a larger fraction of the population is to reduce the cost, size and the difficulty of the operation of the treatment facilities~\cite{Datta2019}. After the accelerator, the next most important component of such a facility is the beam delivery system. For advanced, multidirectional treatment, the beam is transported to the patient by a rotating gantry. The state of the art pencil beam scanning gantries are the PSI Gantry 3 \cite{Koschik:2015exq} in Switzerland, the HIT gantry \cite{Cee:2011zb} in Germany and at the HIMAC facility \cite{Iwata:IPAC2018-TUZGBF1} in Japan. A typical diameter is 10-11\,m for a 100 tons proton gantry and 13\,m for a 600 tons carbon gantry.

The limitations of existing gantries are also related to the maximal energy modulation rate. This is determined by the slow ramping speed of traditional magnets. Scanning is typically performed in steps of 1\% change in beam momentum which has to be done fast enough to achieve short treatment times. Current systems, such as Gantry 2 at PSI, achieve similar range modulation within 80\,ms \cite{CYRSP60}.

One of the requirements of future gantry designs is to accommodate much faster energy changes in the context of linear proton accelerators being currently developed with repetition rates of 200\,Hz \cite{Degiovanni:IPAC2018-MOPML014}. A possible solution is to make use of fixed-field alternating gradient (FFA) optics. Previous work has been done by several groups towards a FFA gantry. Their compactness and operational benefits has motivated initial designs to start from FFA accelerator rings \cite{hud21167}. Some plans for carbon/proton facilities propose large momentum acceptance of $\pm$30\% using tightly-packed high-gradient superconducting magnets \cite{Trbojevic:2010zza}. Another compact approach is based on an extension of the canted cosine theta magnets. Novel combined function sections \cite{Wan:2015zha} are locally achromatic and allow for large momentum acceptance.  

The numerical study presented here investigates the performance of a gantry built on a new concept of adiabatic transition. This technique was firstly used to match the orbit functions in muon acceleration \cite{1288938} over the full energy range by slowly varying the cell parameters between the arc and straight section cells. The goal of the beam-transport line optimised in this study is to deliver the beam at the same position, on the axis, and to provide the necessary spot-size and divergence for all the operating energies. The possible designs examined here offer a more in-depth study of such transitions. A limit on the size of the gantry is obtained based on the performance required for proton therapy~\cite{ffag_ht} and matching of the entire energy range. 

The report is organised as follows: Section 2 introduces the theoretical concepts used to characterise the proton beam, the code used for particle tracking and separate studies on the transition function, end and arc cells; Section 3 describes the methodology for designing a gantry based on adiabatic transition, including the matching section and the size limits imposed by the transition; Section 4 summarises the results and their importance in the light of possible future work and applications to industry.

\section{Preliminary studies}

\subsection{Transverse motion}

During the design phase of an accelerator or a transport line a considerable amount of calculation is centred around the transverse focusing system \cite{Wilson:513326}. A pattern of bending and focusing magnets forms the lattice which determines the size of the beam at each location and consequently, the size and the aperture of the magnets. The repeating unit of a lattice is called a `cell' and it is usually composed of alternating focusing (F) and defocusing (D) quadrupoles. 

\begin{figure}[h]
    \centering
    \includegraphics[width=1\columnwidth]{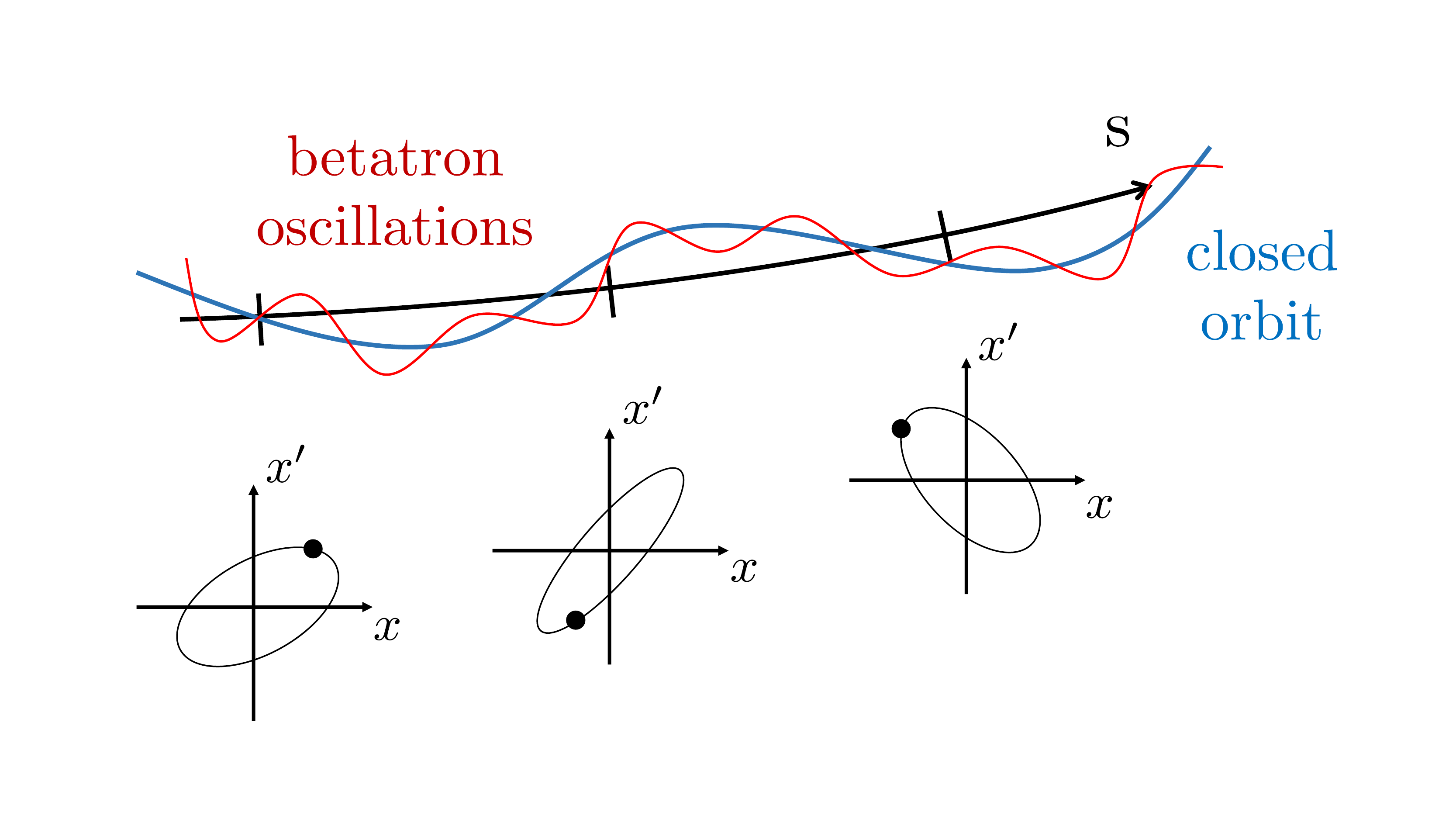}
    \caption{Particle's trajectory as betatron oscillations around the closed orbit and phase space evolution.}
    \label{fig:drawing}
\end{figure}

Passing through this pattern, a charged particle experiences vertical (y) and horizontal (x) transverse oscillations (Fig. \ref{fig:drawing}) described by Hill's equations 
\begin{equation}
    x''+ \left[ \frac{1}{\rho(s)^2} -k(s) \right] x=0
\end{equation}
\noindent
where $\rho(s)$ is the bending radius and $k(s)$ is a periodic coefficient determined by the focusing strength along the line. The solution is similar to the simple harmonic motion
\begin{equation}
    x(s) = \sqrt{\beta_y(s) \epsilon} \cos{[\phi (s) + \phi_0]}
\end{equation}
\noindent
The betatron functions $\beta_x(s), \, \beta_y(s)$ and the betatron phase $\phi(s) = \int ds/ \beta$ have the same periodicity as the lattice; $\epsilon$ is the beam emittance. 

A method to compactly describe the complete dynamics at each point is to use the Twiss parameters $\beta, \alpha, \gamma$. In phase space $(x, x')$ particles travel on ellipses defined by the Courant-Snyder invariant 
\begin{equation}
    \gamma(s)x^2 + 2 \alpha(s) x x' + \beta(s) x'^2 = \epsilon
\end{equation}
This ellipse has an area $\pi \epsilon$. The values of $\epsilon$ and $\beta(s)$ are a measure of the beam size as the maximum amplitude of the betatron motion is $\sqrt{\beta(s) \epsilon}$ \footnote{The complete orbit of each particle is given by the amplitude of oscillation around the closed orbit and the deviation of the closed orbit due to dispersion (\ref{eq:disp}) - Fig. \ref{fig:drawing}.}. In transport lines, the Twiss parameters are propagated along the lattice with the help of transfer matrices for each element.  

In a cell with bending fields, the stable closed orbit solution for the motion of particles is matched to some ideal momentum $p_0$. Small changes in momentum $\Delta p$ will cause particles to follow a slightly different orbit. The change in the radius of the new orbit $\Delta \rho$ is defined by the dispersion D(s)

\begin{equation}
    \label{eq:disp}
    \Delta \rho = D(s) \frac{\Delta p}{p_0}
\end{equation}

\subsection{FFA}

The fixed field alternating gradient (FFA) accelerators~\cite{ffa1,Symon1956} combine the early principles of two other machines: the constant field magnets of the cyclotron and the strong focusing present in the synchrotron \cite{COURANT19581}. Their potential applications include the production of high-power beams~\cite{MORI2000300,ruggiero2005ffag}, hadron therapy~\cite{Keil2007,ffag_ht,osti_1573824}, and muon acceleration~\cite{Summers2005,PLANCHE201021}. 

There are two types of FFA accelerators: scaling and non-scaling. For the former variety the shape of the orbits is the same while their size scales with energy. For the latter type, the orbit scaling condition is relaxed making the orbits much more compact over a large momentum range \cite{Sheehy:2016kvm}. However, due to unequal focusing strengths for different orbits, the number of particle oscillations in one period vary with energy. Hence, during the accelerating cycle, significant beam degradation is caused by resonance crossings, unless the acceleration rate is sufficiently high \cite{baartman2004fast}.

The functioning of such a device was first demonstrated by the Electron Machine for Many Applications (EMMA) \cite{Machida2012}, a linear non-scaling FFA which uses only dipolar and quadrupolar field components. By analogy, a FFA transport-line 
is made of fixed-field combined function magnets with linear transverse field gradients. Since this is a single-pass system the resonance crossing is not an issue. 
\subsection{Adiabatic transition}

One of the requirements of the gantry is to bring all the beams with different energy in the operating range onto the same point at the patient. Multiple transition sections are employed in the design presented here to bring the orbits onto or close to the axis of the straight cells. A slow spatial variation of the cell parameters allows the orbits to adiabatically approach the periodic orbit solutions inside each cell. 

The methodology followed in this study was developed and tested as a transition between straight and arc sections of a return loop for the CBETA energy recovery linac \cite{Hoffstaetter:2017jei}. At cell $i$ in the transition, each parameter is set by sampling from a smooth function $f_T$ according to 

\begin{equation}
    p_i = \left[ 1 - f_T \left( \frac{i}{n_T+1} \right) \right] p_{init} + f_T \left(\frac{i}{n_T+1} \right) p_{final}
\end{equation}
where $n_T$ is the total number of cells in the transition. The function $f_T$ is determined by an arbitrary number of coefficients $a_k$ as

\begin{equation}
    f_T(x) = \frac{1}{2} + \left(x - \frac{1}{2} \right) \sum_{k=0} a_k \binom{2k}{k} x^k (1-x)^k
\end{equation}
where $a_0 = 1$ to ensure that $f_T(0) = 0$ and $f_T(1) = 1$. The coefficients $a_1, a_2, a_3$ were tuned sing a transition made of ten cells from an arc configuration to a straight section. Through the transition, each cell has a smoothly varying bend angle and dipole field.

\begin{figure}
    \centering
    \hspace*{-0.8em}
    \includegraphics[width=1.04\columnwidth]{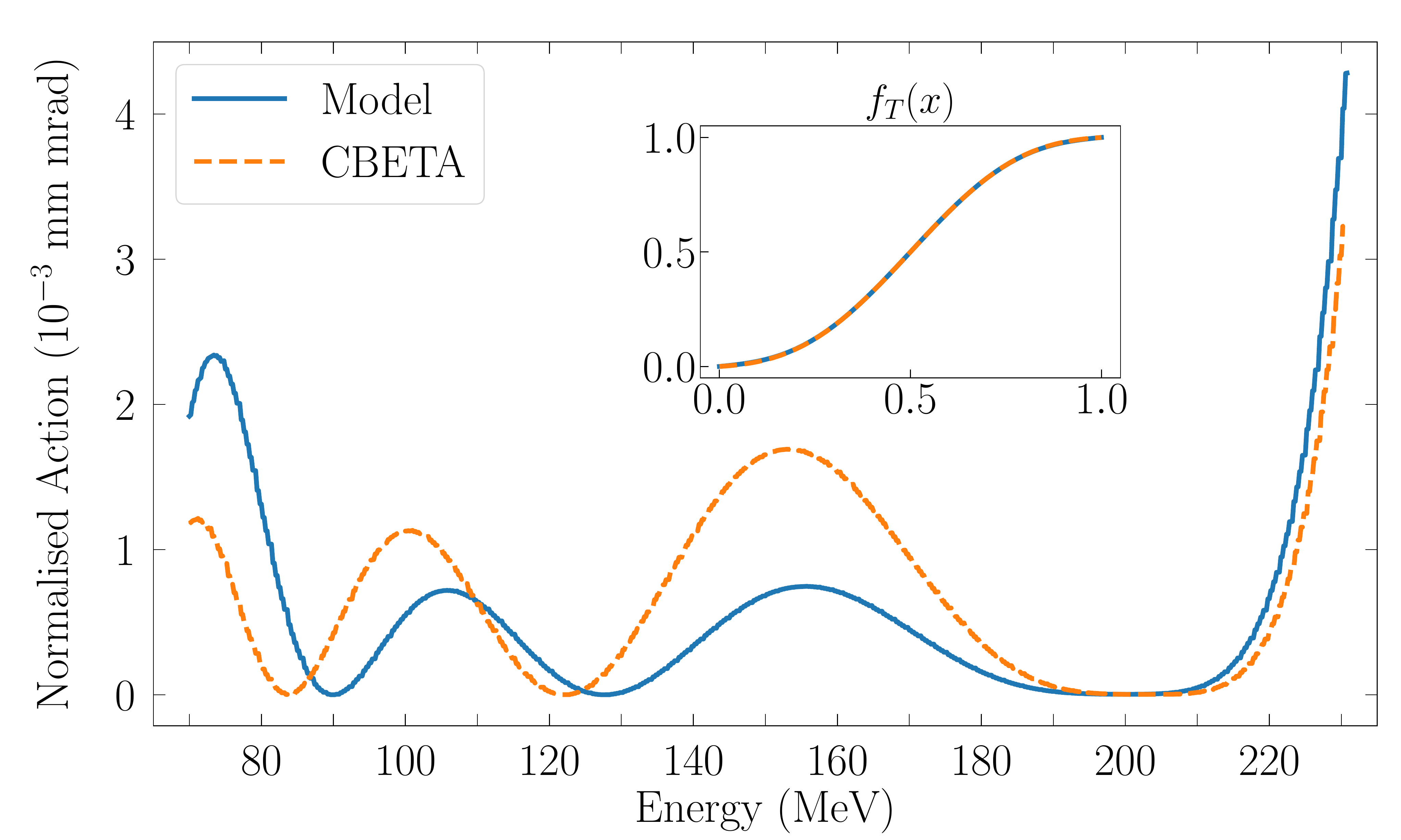}
    \caption{$J_{norm}(E)$ in the x-plane corresponding to the transition $f_T(x)$. 'Model' is the result of our optimisation. `CBETA' corresponds to the coefficients from \cite{Berg2017}.}
    \label{fig:trans_func}
\end{figure}

The deviations from an ideal orbit matching is measured using the normalised action~\cite{Hoffstaetter:2017jei} 
\begin{equation}
\label{eq:J_norm}
     J_{norm}(E) = \frac{p}{2m_pc} \left( \gamma_x x^2 + 2 \alpha_x x p_x + \beta_x p_x^2 \right)
\end{equation} 
 \noindent
 at the straight cell, where $p_x$ is the normalised phase space momentum. $J_{norm}$ gives an approximation to the emittance growth. Firstly, a particle on the periodic orbit in the arc cell is tracked through the transition. Secondly, the normalised action of the test particle is determined at the straight cell, treating the straight cell as periodic. Table \ref{table:trans_func} shows the result of our optimisation model in comparison to that of \cite{Berg2017}.
 
 A further comparison is presented in Fig. \ref{fig:trans_func}. While the overall shape of the transition $f_T(x)$ is not changed significantly, small increases in the slope near $x=0$ and $x=1$ determine an increase in $J_{norm}(E)$ at lower energies. For the ten cell transition used here, smaller deviations from the periodic orbit were obtained in the energy range 90--210\,MeV, but an increase by a factor of about 2 was observed for the lower energies 70--90\,MeV.

\begin{table}
\caption{Coeff. $a_k$ used in the transition $f_T(x)$}
\label{table:trans_func}
\vspace{-0.75cm}
\begin{center}
\begin{tabularx}{\columnwidth}{c l l l}
\hline
Model  & $a_1$: 0.835 & $a_2$: 0.813 & $a_3$: 0.237 \\
CBETA$^{[8]}$ & $a_1$: 0.894 & $a_2$: 0.659 & $a_3$: 0.329 \\
\hline
\end{tabularx}
\end{center}
\end{table}

\subsection{BMAD}

The \texttt{Bmad} code in parallel with the \texttt{Tao} program were used to model the components of the gantry, run the optimisation procedures and analyse the performance of the beam-transport. \texttt{Bmad} is a software library \cite{Sagan:2006sy} for relativistic charged-particle simulations in high energy accelerators and storage rings, developed at Cornell University\footnote{The \texttt{Bmad} Manual can be obtained at: www.classe.cornell.edu/bmad}. It cuts down the time required to build new types of simulations and it reduces the programming errors. 

To find the particle orbits, the code calculates the transfer matrices and Twiss parameters at each lattice element specified in the initialisation file. Each type of element has several tracking methods available. We have used symplectic tracking based on a Hamiltonian with Lie operators \cite{Forest:1998th}. 

\texttt{Tao} (Tool for Accelerator Optics) is a general purpose program, based upon \texttt{Bmad}. It contains routines to view and optimise lattices, do Twiss and orbit calculations and track single or multiple particles. The main advantage of these simulation tools is that they include a proper treatment of the FFA magnets. The implemented theoretical models allow for precise calculations even when the momentum deviation is large. 

Several algorithms are available as optimisers. They vary the lattice model in order to minimise a merit function based on the variables defined by the user. During the cell parameter search, we used a global optimiser in combination with a local optimiser one after the other. The algorithm used for global search is built on differential evolution \cite{542711}. Additional care was taken during each optimisation for finding an appropriate value of the step size. A larger step size increases the parameter space being explored, but it makes it harder to find the local minimum. The local search algorithm calculates the derivative matrix and takes steps in variable space accordingly.  

\begin{figure}
    \centering
    \includegraphics[width=1\columnwidth]{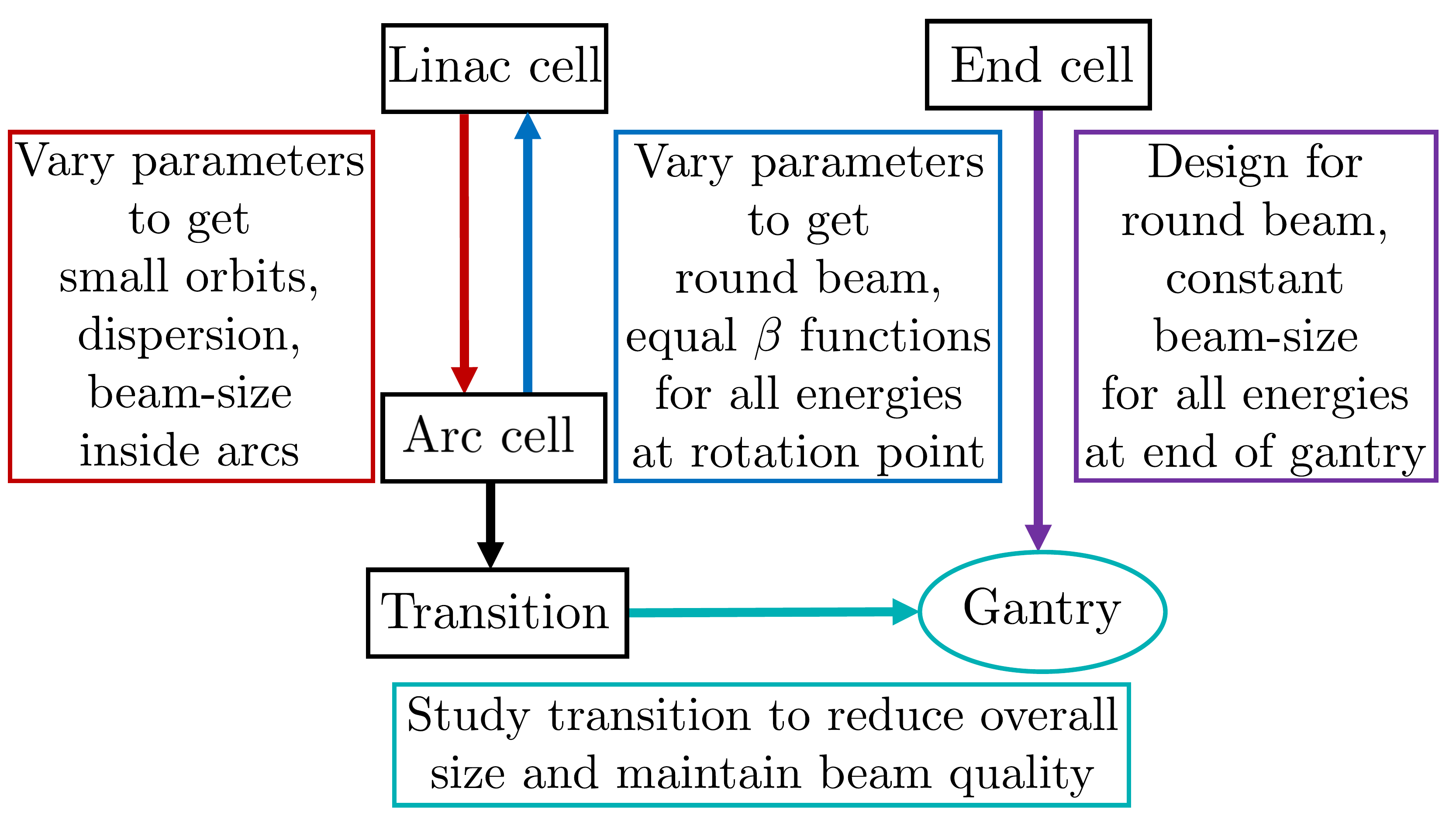}
    \caption{Schematic representation of the optimisation strategy.}
    \label{fig:flow}
\end{figure}

Fig. \ref{fig:flow} shows a schematic summary of the strategy towards the overall optimisation process. The starting point is either the cell of the linac or the arc cell. 

\subsection{ADAM linac cell}
\label{ADAM_linac}

The gantry proposed in this study is designed to deliver a beam from a linear accelerator. In the past few years, the CERN spin-off company ADAM (Application of Detector and Accelerators to Medicine) has developed a prototype for the LIGHT accelerator -- the first commercial linac for proton therapy~\cite{Degiovanni:IPAC2018-MOPML014}. Its features make it well suited for the active spot scanning technique. The beam structure is pulsed with electronic modulation of energy and intensity in 5\,ms. The high frequency of the injection section (the RFQ) allows an extremely small beam emittance, $\epsilon_n = 0.25 \, \textrm{mm mrad}$. Hence, the beam size is reduced and magnets with small aperture can be used for the transfer lines and the gantry. 

The beam can be accelerated up to 230\,MeV with almost no losses using a modular approach. The final section of Cell Coupled Linac (CCL) structures is based on the design of the LIBO prototype \cite{Amaldi_ugo}, built and tested as a proof of principle in 2000. The narrow beam is focused by small Permanent Magnetic Quadrupoles (PMQs) arranged in a FODO lattice between the accelerating tanks. The lower energy limit of 70 MeV is given by the physical aperture of the quadrupoles as the physical emittance becomes larger as the energy decreases. 

In order to match the beam optics, a model was created for the last cell in the linac similar to the modules of the LIBO-62 design \cite{AMALDI2004512}. The FODO cell has equal drift spaces and the specifications from Table \ref{table:linac_cell} in the appendix. The cell parameters were chosen to ensure stability and a round beam at the end with $\beta_{x,y} = 1.2 \, \textrm{m}$ for the entire energy range.

\subsection{The base cell}
\label{Basic_cell}

The beam-line is built out of repeating cells with the same overall structure. Each cell is composed of two non-scaling FFA permanent magnets: a focusing quadrupole and a combined function dipole with a defocusing component.

The large momentum acceptance of this structure comes from the strong focusing (large gradients, short drift spaces), and from the very small dispersion. Optimal lattice designs for small dispersion were investigated in the context of synchrotron light sources \cite{low_emitt} using the normalised dispersion $\mathcal{H}$, defined as
\begin{align}
    \mathcal{H} &= X_d^2 + P_d^2 \\
    X_d &= D / \sqrt{\beta_x} \\
    P_d &= \left( \alpha_x D + \beta_x D' \right) / \sqrt{\beta_x} 
\end{align}
\noindent
where $D$ and $D'$ are the dispersion function and its derivative, $\beta_x$ and $\alpha_x$ are the horizontal Twiss parameters. $\mathcal{H}$ is invariant in regions with no dipoles \cite{Lee}. Across a thin dipole with the bending angle $\theta$, the above functions evolve as
\begin{align}
    \Delta X_d  &= 0 \\
    \Delta P_d &= \sqrt{\beta_x} \Delta D' = \sqrt{\beta_x} \theta
\end{align}
A minimum dispersion function is achieved by having the minimum of the horizontal amplitude $\beta_x$ at the centre of the bending element~\cite{Trbojevic2003FFAGLF}. Thus, a bending component is added to the focusing quadrupole where the $\beta_x$ function is minimum. 

\begin{figure}
    \centering
    \hspace*{-1em}
    \includegraphics[width=1.1\columnwidth]{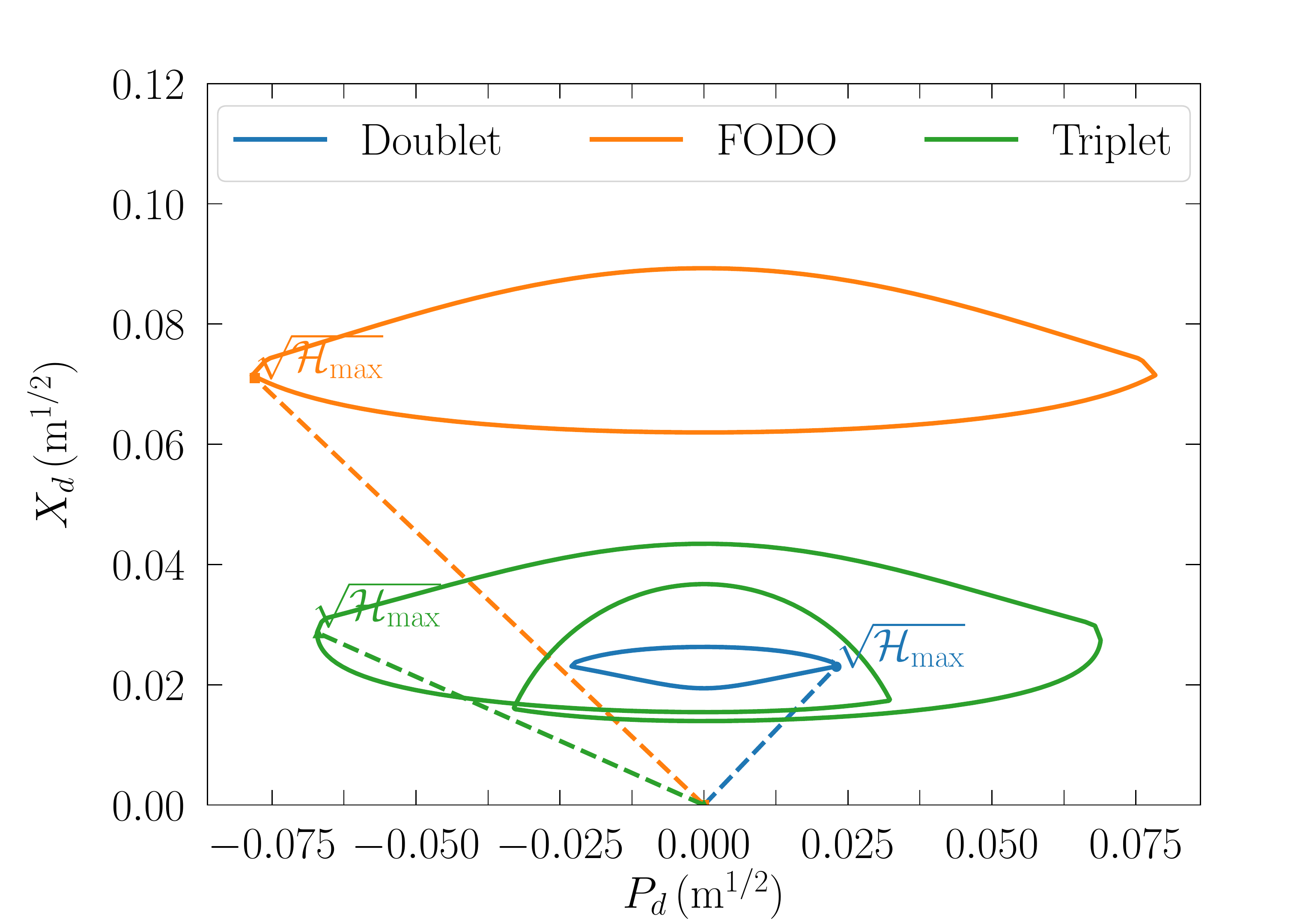}
    \caption{Normalised dispersion space and the function $\mathcal{H}$ for three cell configurations (using the same magnets).}
    \label{fig:curly-H}
\end{figure}

An analysis of three basic cell configurations is shown in Fig.~\ref{fig:curly-H}. A doublet structure as that described at the beginning of the subsection provides a smaller $\mathcal{H}$-function over the entire cell and, hence, smaller maximum dispersion.  

A further important aspect during the design of the cell is the choice of magnets. While permanent magnets (PM) were developed mainly for insertion devices on light sources, there has been a growing interest in recent years for PM-based beam-lines. These are attractive in terms of running costs and stability. 

\begin{figure}[t!]
    \centering
    \includegraphics[width=0.9\columnwidth]{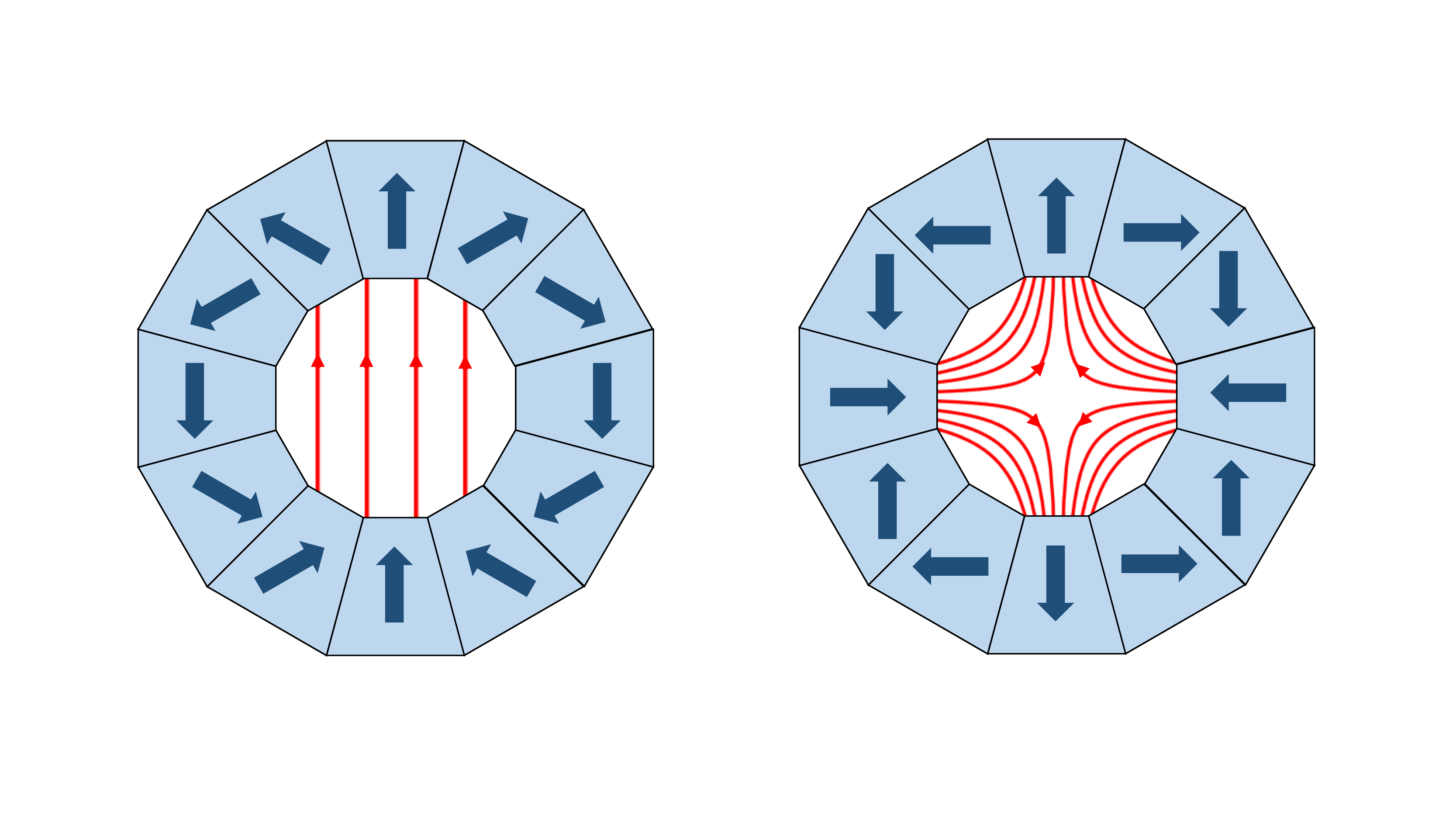}
    \caption{Geometry of the Halbach magnets (dipole and quadrupole fields).}
    \label{fig:Halbach_mag}
\end{figure}

Our design is based on Halbach arrays~\cite{HALBACH19801} which generate multipole fields from permanent magnetic blocks with varying magnetisation angles (Fig.~\ref{fig:Halbach_mag}). They typically have very small aperture and relatively high fields (1.2--1.4\,T) and gradients (up to 170\,T/m) achievable. The size of a quadrupole can be estimated using a simple dependence on the necessary gradient~\cite{Benabderrahmane:IPAC2017-THYB1}

\begin{equation}
\label{eq:Halbach}
    G = 2 B_r K \left( \frac{1}{r_i} - \frac{1}{r_e} \right)
\end{equation}
\noindent
where $K \rightarrow 1$ as the number of segments increases and $B_r \approx 1.3 \, \mathrm{T}$ for recent NdFeB type materials. These magnets require small operating temperatures (around 70$^{\circ}$ C) and they are more sensible to radiation damage. 


In each of the following studies, the aim was to find an optimal arc cell given a set of parameters which were being varied. Each time, the cell was optimised with decreasing weight for:
\begin{itemize}
    \item stable orbits for the entire energy range
    \item minimum orbit excursions for 70\,MeV and 230\,MeV beams
    \item minimum highest values of $\beta$-functions in both planes for 70\,MeV, 140\,MeV, and 230\,MeV beams
    \item minimum total cell length. 
\end{itemize}
Other constraints were considered: minimum gradients, maximum bending angle or equal orbit excursions on both sides of the central axis. However, those listed above were found to give the best results in terms of cell performance. 

\begin{figure}
    \centering
    \hspace*{-0.8em}
    \includegraphics[width=1.05\columnwidth]{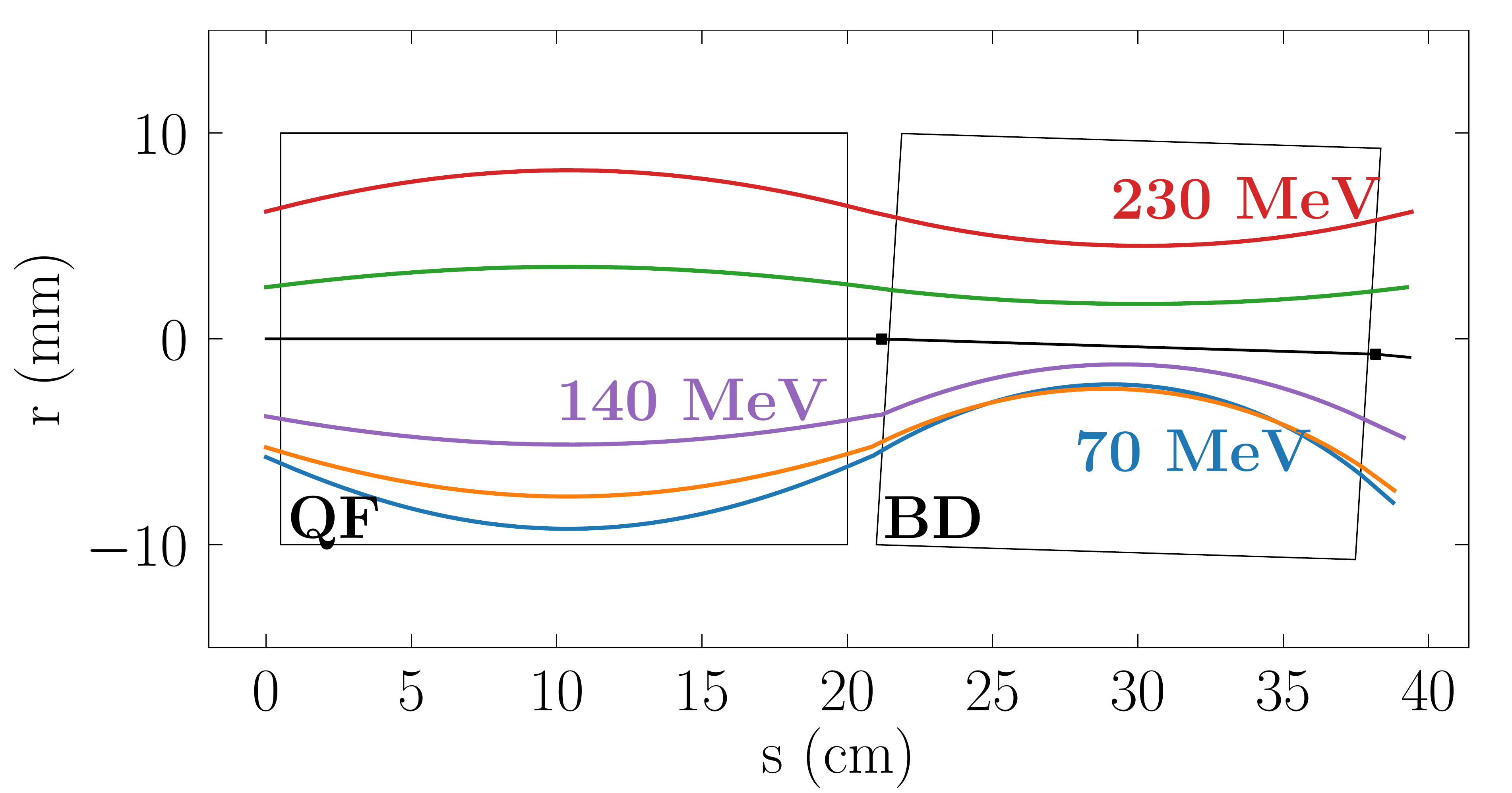}
    \caption{Arc cell and periodic orbits for 5 equally separated momenta, labelled by energy. The black points mark the bending of the beam-pipe. QF -- focusing quadrupole, BD -- bending dipole with a defocusing component.}
    \label{fig:cell_orbits}
\end{figure}

The result of such an optimisation is showed in Fig.~\ref{fig:cell_orbits}. The periodic orbits correspond to the shortest cell with small aperture and small orbit excursions given the values of magnetic field strength and gradient achievable with Halbach permanent magnets. The exact parameters of the cell are included in the appendix.

\begin{figure}
    \centering
    \hspace*{-0.8em}
    \includegraphics[width=1.03\columnwidth]{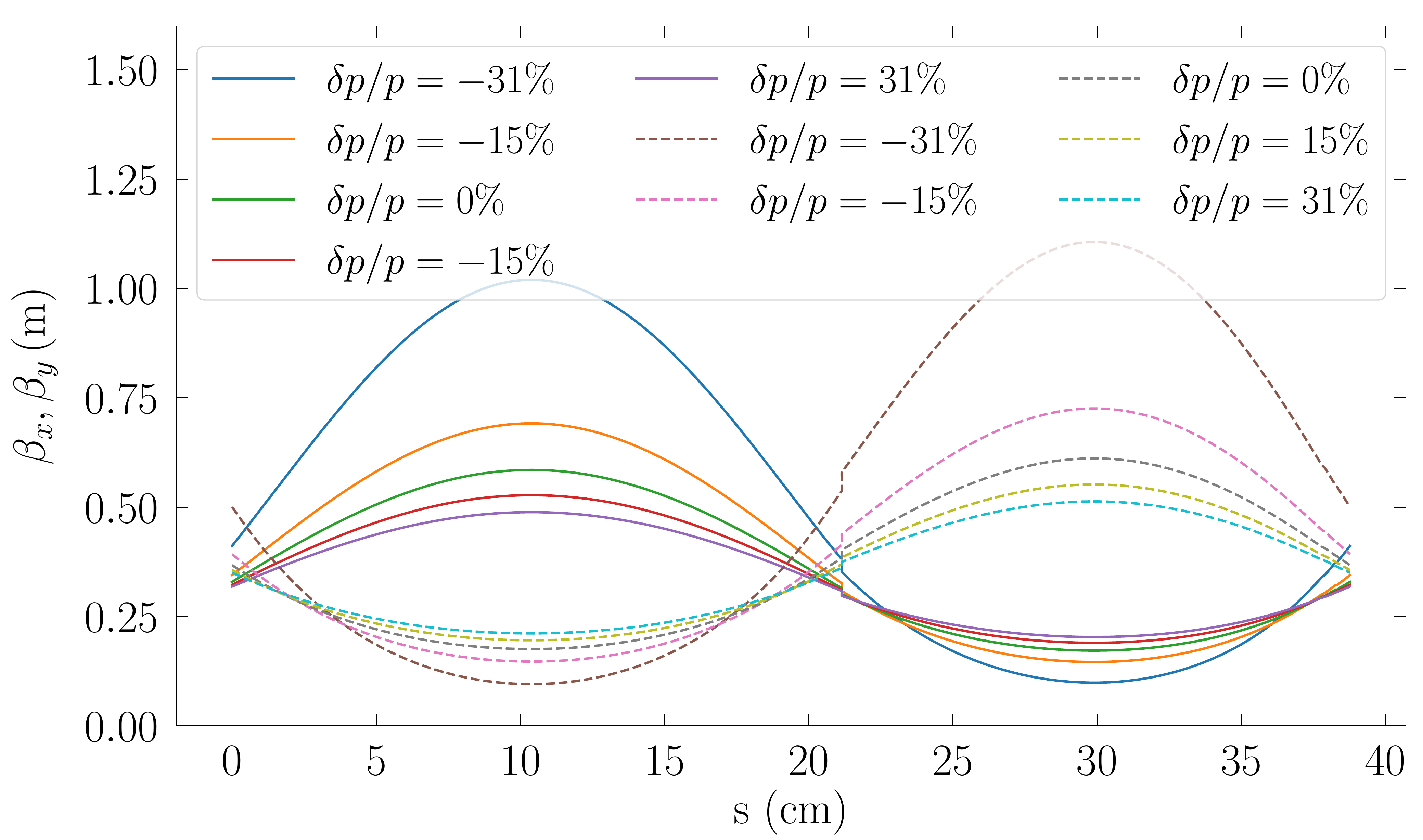}
    \caption[Caption for LOF]{Betatron functions in the arc cell for several momentum deviations. Full lines show $\beta_x$; Dotted lines show $\beta_y$. Edge effects are included in the calculation\protect\footnotemark.}
    \label{fig:betas}
\end{figure}

\footnotetext{The discontinuities at the entrance and exit faces of the bending dipole are caused by the edge focusing effect; particle tracking is done using a hard-edge approximation for the magnets.}

\subsection{End cell}

The cell at the end of the gantry was designed to provide certain characteristics for the beam at the isocentre -- the intersection point between the axis of rotation of the gantry and the treatment couch. The spot is required to be a symmetric, 2D Gaussian shape with FWHM typically between 3 and 10 mm~\cite{Giordanengo:2018tqo}, as required by the spot scanning technique to achieve small depositions of the dose. 

\begin{figure}
    \centering
    \hspace*{-0.8em}
    \includegraphics[width=1.05\columnwidth]{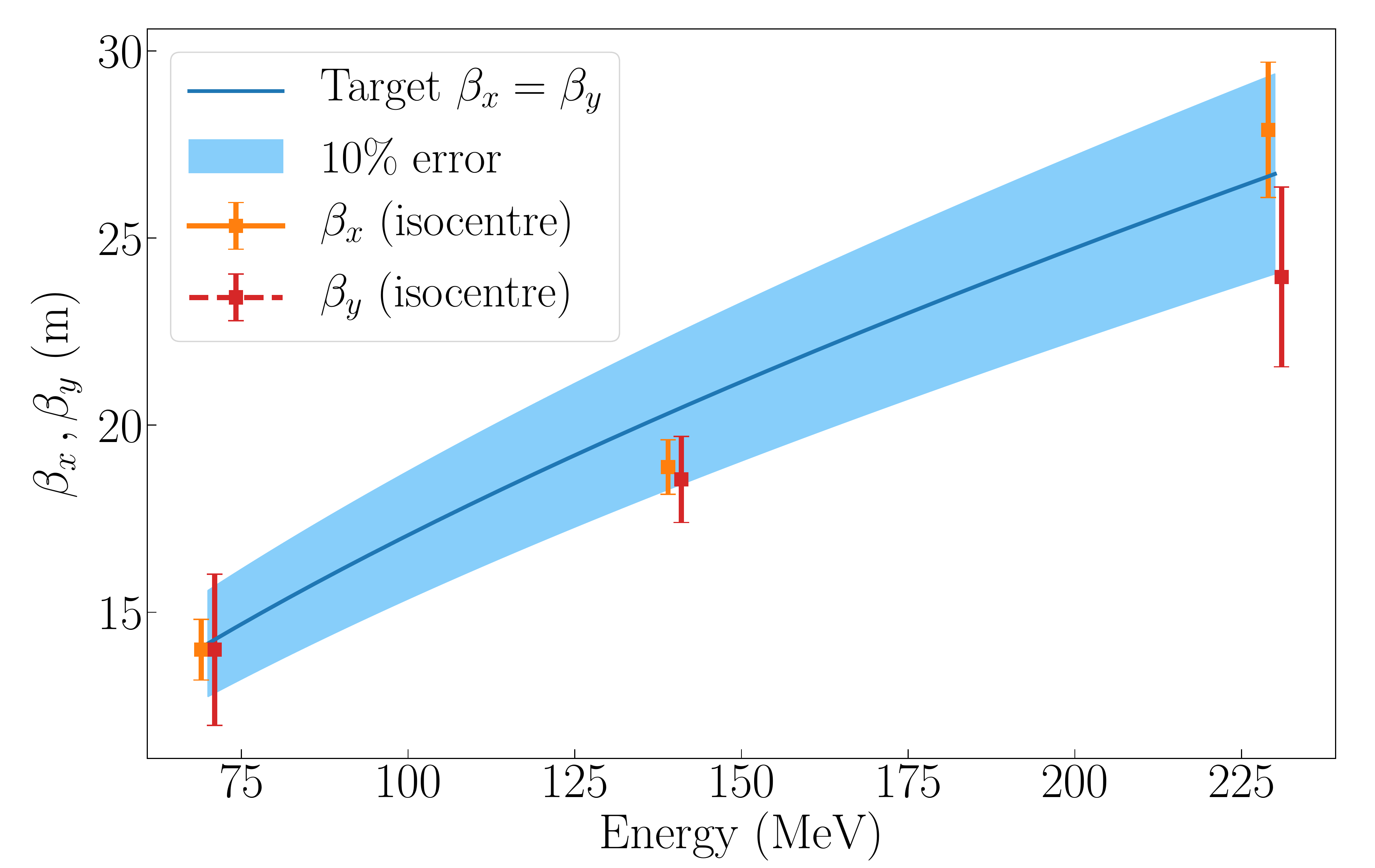}
    \caption{Values of the betatron functions for the optimised cell at the end of the gantry at three representative energies (dots). The target values required to obtain a round beam of constant size at isocentre is indicated by the blue region. The error bars show the variation of the beta function for vertical displacement of $\pm 10$ cm with respect to the isocentre.}
    \label{fig:end_optics}
\end{figure}

For a Gaussian beam, the spot size is given by

\begin{equation}
    \sigma_{x,y} = \sqrt{\epsilon \beta_{x,y}}
\end{equation}
\noindent
where we consider equal geometric emittance $\epsilon_x = \epsilon_y = \epsilon$ in both planes. Due to the adiabatic damping effect \cite{Wolski:2014lba}, the quantity that is conserved is the normalised emittance 

\begin{equation}
    \epsilon_n = \epsilon \beta \gamma
\end{equation}
\noindent
where $\beta \, ,\gamma$ are the usual relativistic factors. Thus, we expect a decrease in $\epsilon$ with increasing energy. An approximately constant beam size can be maintained at at all energies by using the appropriate optics in the end cell. Fig.~\ref{fig:end_optics} shows the necessary $\beta_x, \, \beta_y$ where a 10\% variation is tolerable for treatment. 

\begin{figure}
    \centering
    \hspace*{-0.8em}
    \includegraphics[width=1.04\columnwidth]{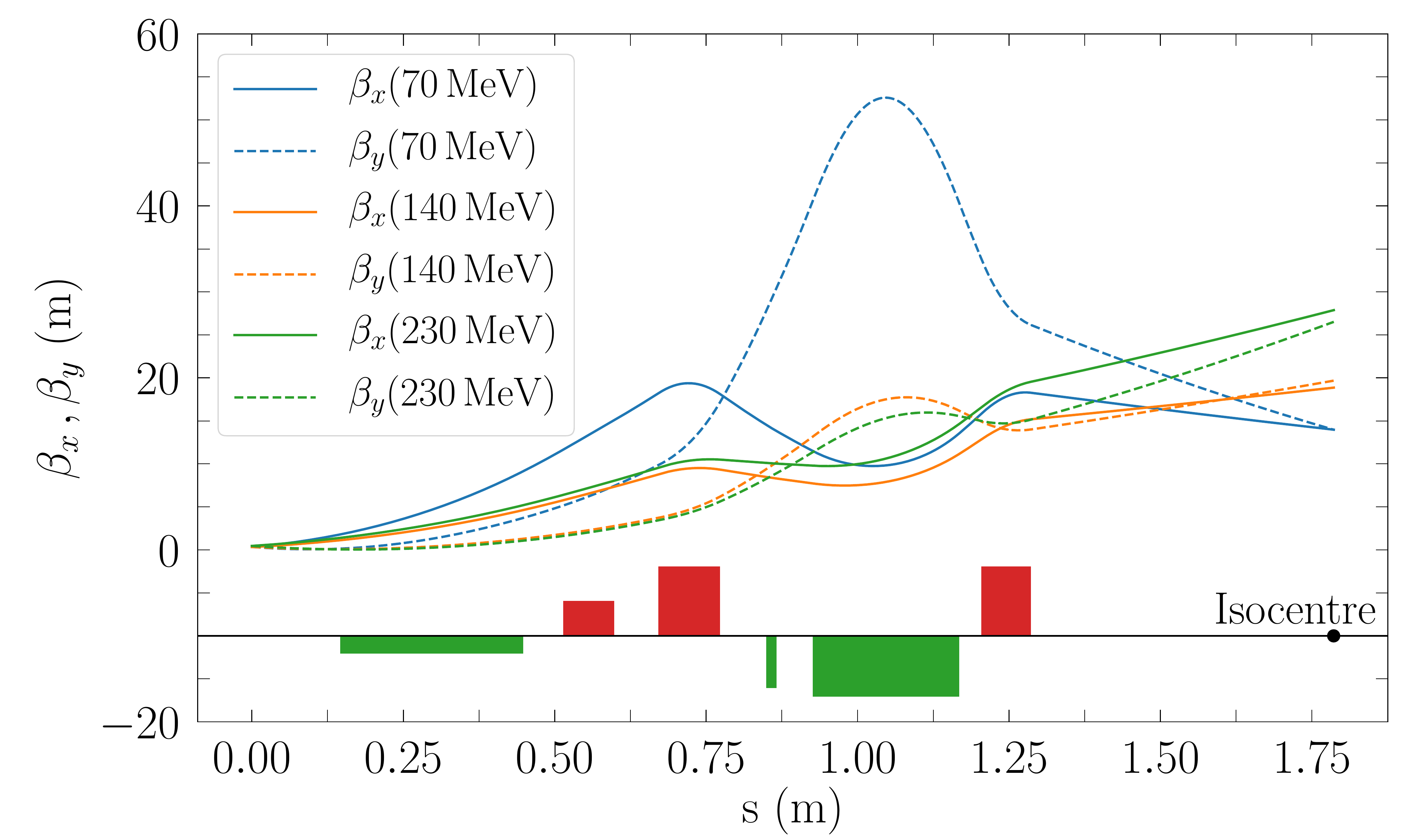}
    \caption{Illustration of the cell at the end of the gantry (green -- defocusing quadrupole, red -- focusing quadrupole) and $\beta$-functions at the isocentre. The relative heights of the magnets represent their relative strengths.}
    \label{fig:end_cell}
\end{figure}

To design the cell at the end of the gantry, two set of conditions were imposed at the isocentre for three energies (70 MeV, 140 MeV and 230 MeV). Firstly, the beam was required to be round and the values of the betatron functions to correspond to a beam size of $\sigma = 3 \, \mathrm{mm}$ and normalised emittance $\epsilon_n = 0.25 \, \mathrm{mm \, mrad}$. Secondly, the slopes $\alpha_x,\alpha_y$ were minimised to obtain parallel beams with small diameter change around the isocentre since it simplifies the treatment planning. An illustration of the result is shown in Fig.~\ref{fig:end_cell}. The distance from the last gantry element to the patient is rather small and offers limited space for placing the necessary beam position and dose monitors. However, Fig.~\ref{fig:end_cell} demonstrates that a suitable beam optics can be found.

\section{Gantry optimisation}
\subsection{Gantry layout}

\begin{figure}
    \centering
    \includegraphics[width=1\columnwidth]{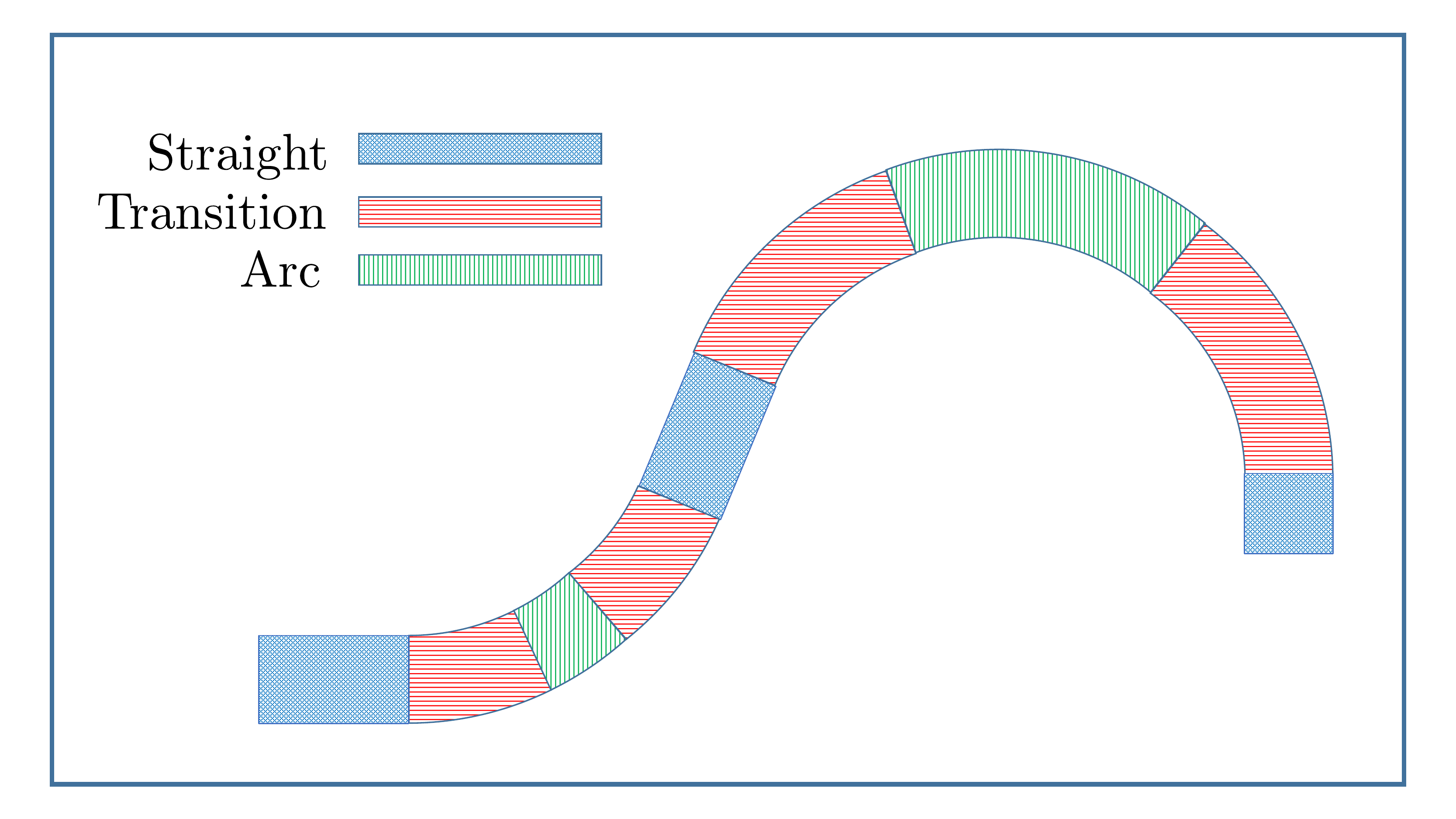}
    \caption{Gantry layout starting at the last straight cell of the linac and the three types of sections used.}
    \label{fig:layout}
\end{figure}

A symmetric design, schematically shown in Fig.~\ref{fig:layout}, was chosen to compensate for the build-up of matching errors and for the increase of dispersion in the transition sections. At the middle of the arc sections, the layout of the cells is reversed. Thus, the gantry is formed by two regions of the type straight-arc-straight with four transition segments. 

One of the optics requirements for the gantry is zero dispersion at the points where the sign of the curvature changes. This is achieved by introducing a straight cell at the inflection point. The number of cells in the arc sections can be varied to change the distance from the end of the gantry to the patient.

\subsection{Study of the transition} 

The gantry is designed to be used in combination with a linac such as that described in Sec. \ref{ADAM_linac}. Based on the optics in the linac, the performance of the gantry was analysed when the first adiabatic transition starts directly from the last cell of the linac. Starting from the last cell of the linac, four combinations of parameters were varied in the transition sections as follows:
\begin{enumerate}[label=(\Alph*)]
    \item bending angle and dipole field
    \item bending angle, dipole field and drift spaces
    \item bending angle, dipole field, drift spaces and focusing gradients
    \item all parameters of a cell
\end{enumerate}

\begin{figure*}[h]
    \centering
    \hspace*{-1em}
    \includegraphics[width=1.1\textwidth]{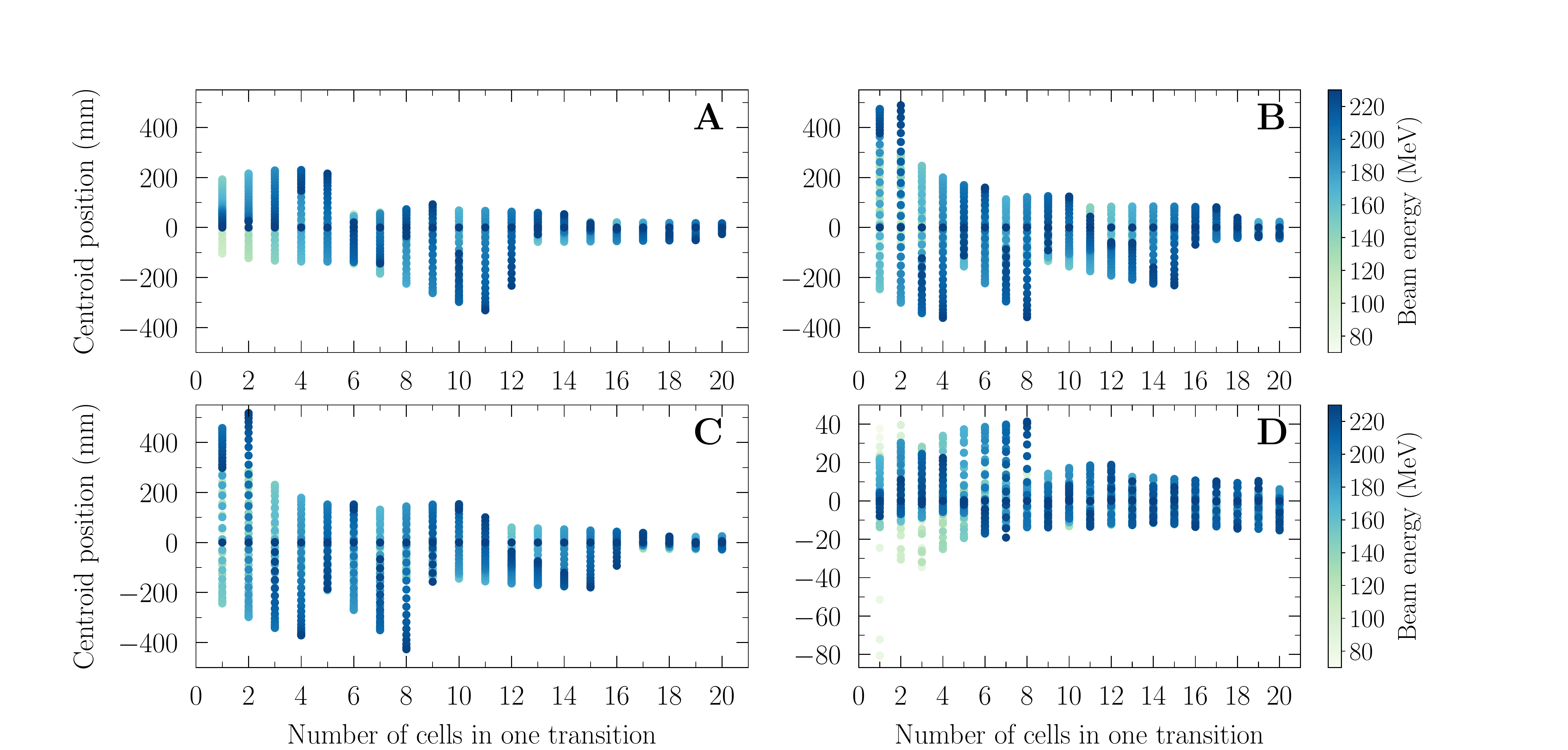}
    \caption{The position of the beam centroid (x-plane) at the end of each gantry. A, B, C, D represents four separate sets of parameters being varied in the transition sections. The scale of the orbit oscillations for set D is one order of magnitude smaller.}
    \label{fig:orbit_analysis}
\end{figure*}

\noindent
To measure the efficiency, a Gaussian beam of 1000 particles was tracked through the gantry and the centroid position was calculated at the end. The centroid corresponds to the mean of the particle distribution at the end when a Gaussian is fitted along the two axes.

As can be seen in Fig.~\ref{fig:orbit_analysis}, for the length of the transition sections being fixed, the centre of the beam oscillates around the axis when the energy is changed. The parameter set D gives orbit deviations smaller than the other sets by a factor of 10. Even in this case, the beam position is comparable to the size of the orbit of single particles and has the same dimensions as the aperture of the magnets. 

Consequently, a beam cannot be transported to the central axis at the end of the gantry for all the necessary energies even by using up to twenty cells in one transition. A different cell is required at the start the gantry. Due to the space required for the accelerating cavities, a cell designed for a linac has short quadrupoles and long drift spaces relative to the total length of the cell. By contrast, an arc cell suitable for a transport line has short drift spaces and long quadrupoles to maintain a small beam size and small orbit excursions. 

However, a solution was found based on the reverse approach. In the final design, the optimisation process started from a suitable arc cell. The base cell was optimised as described in Sec.~\ref{Basic_cell} for small magnet apertures. Then, the transition is kept to the minimal set of changing parameters, varying only the bending angle and the bending field to obtain a suitable straight cell at the start of the gantry.

\subsection{Matching section} 

Since the last cell of the linac and the first cell of the gantry are now different, an additional matching section is required. Nonetheless, this transport line is usually a requirement of therapy facilities since the accelerator complex and the gantry are some distance apart.

\begin{figure}[t!]
    \centering
    \hspace*{-0.8em}
    \includegraphics[width=1.05\columnwidth]{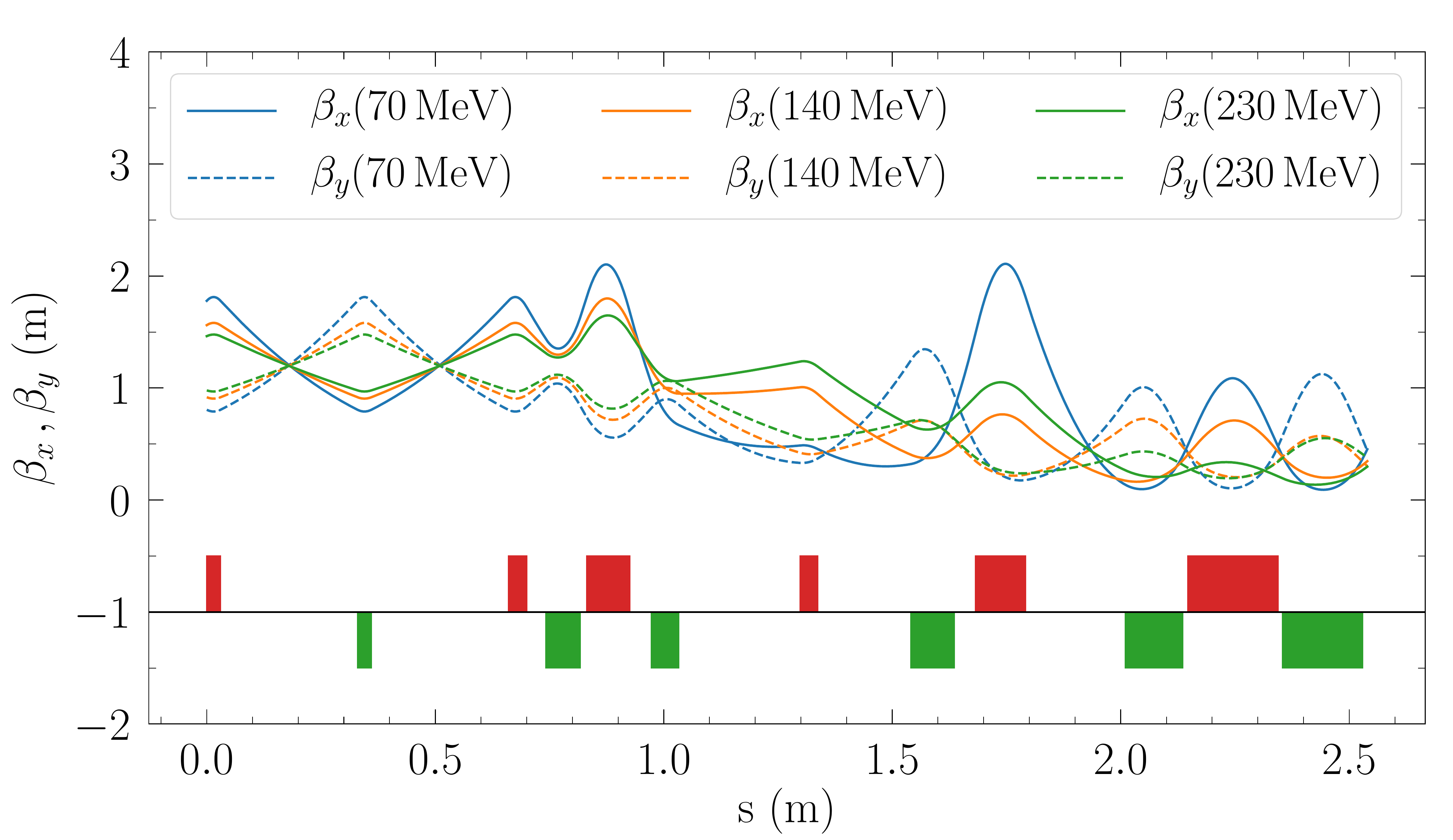}
    \caption{The last cell of the linac followed by a matching section (layout and optics). The energies shown are those used in the matching.}
    \label{fig:Trans_line}
\end{figure}

Fig.~\ref{fig:Trans_line} shows a matching section composed of five cells. An adiabatic matching section does not work in this case. To keep the beam size for lower energies at reasonable values, a length of about 2.5\,m is required. Matching was achieved for 70\,Mev, 140\,MeV, and 230\,MeV. As the $\beta$-functions are approximately equal at the end for all the energies, the errors are small for the rest of the energy range. The extension of the transport line is possible at both ends by simply repeating the first or the last cell.

\subsection{Final design of the gantry}

\begin{figure}
    \centering
    \vspace{-0.9em}
    \hspace*{-1em}
    \includegraphics[width=1.1\columnwidth]{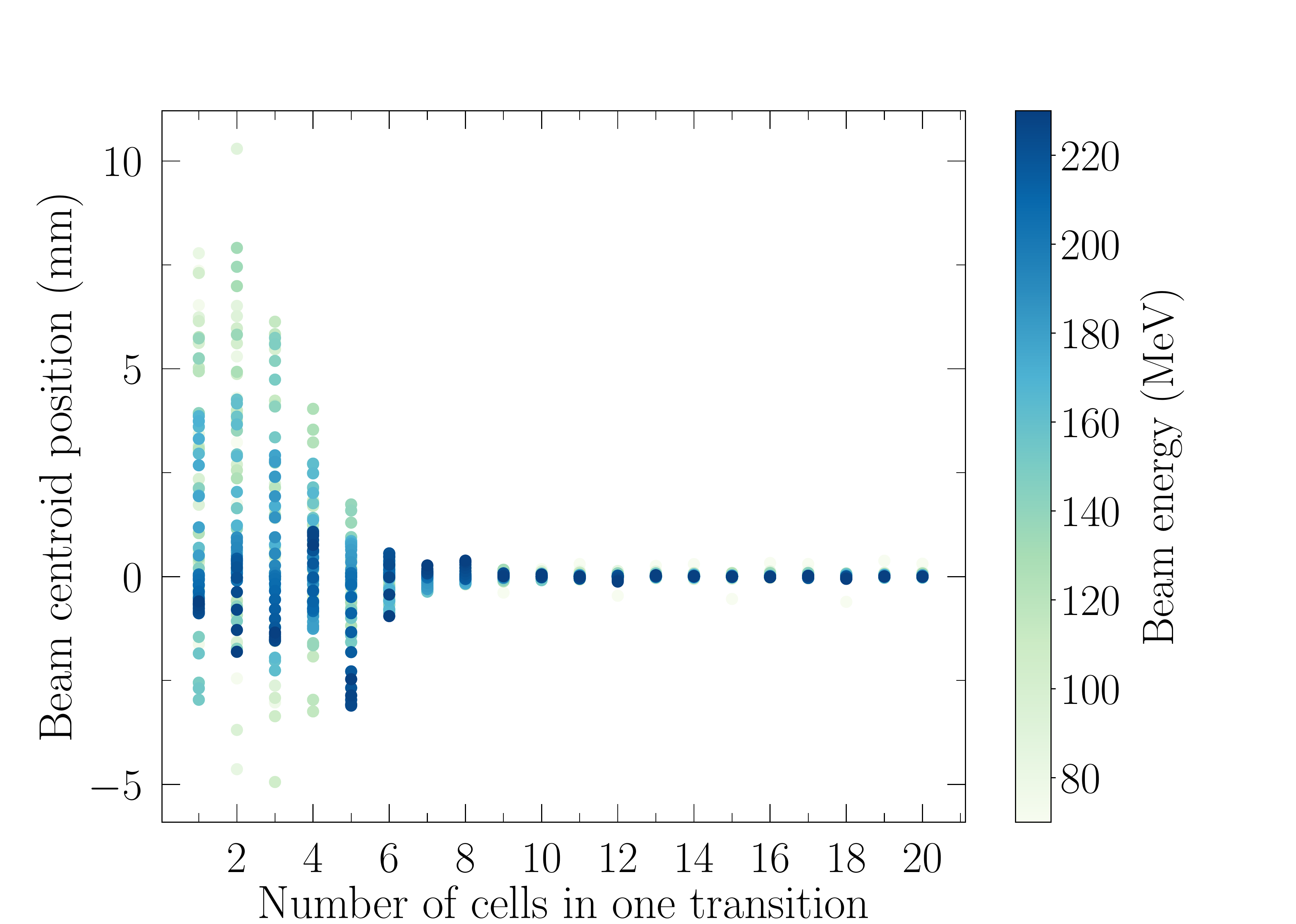}
    \caption{The position of the beam centroid (x-plane) at the end of the proposed gantry.}
    \label{fig:FG_beam_orbit}
\end{figure}

An analysis of the position of the beam centroid at the end of the gantry was used to set the length of the transition sections. Fig.~\ref{fig:FG_beam_orbit} shows that, when a beam is tracked through the gantry, the beam centroid remains near the axis for all energies if the number of cells in one transition section is higher or equal to seven. 

\begin{figure}
    \centering
    \hspace*{-0.6em}
    \includegraphics[width=1.05\columnwidth]{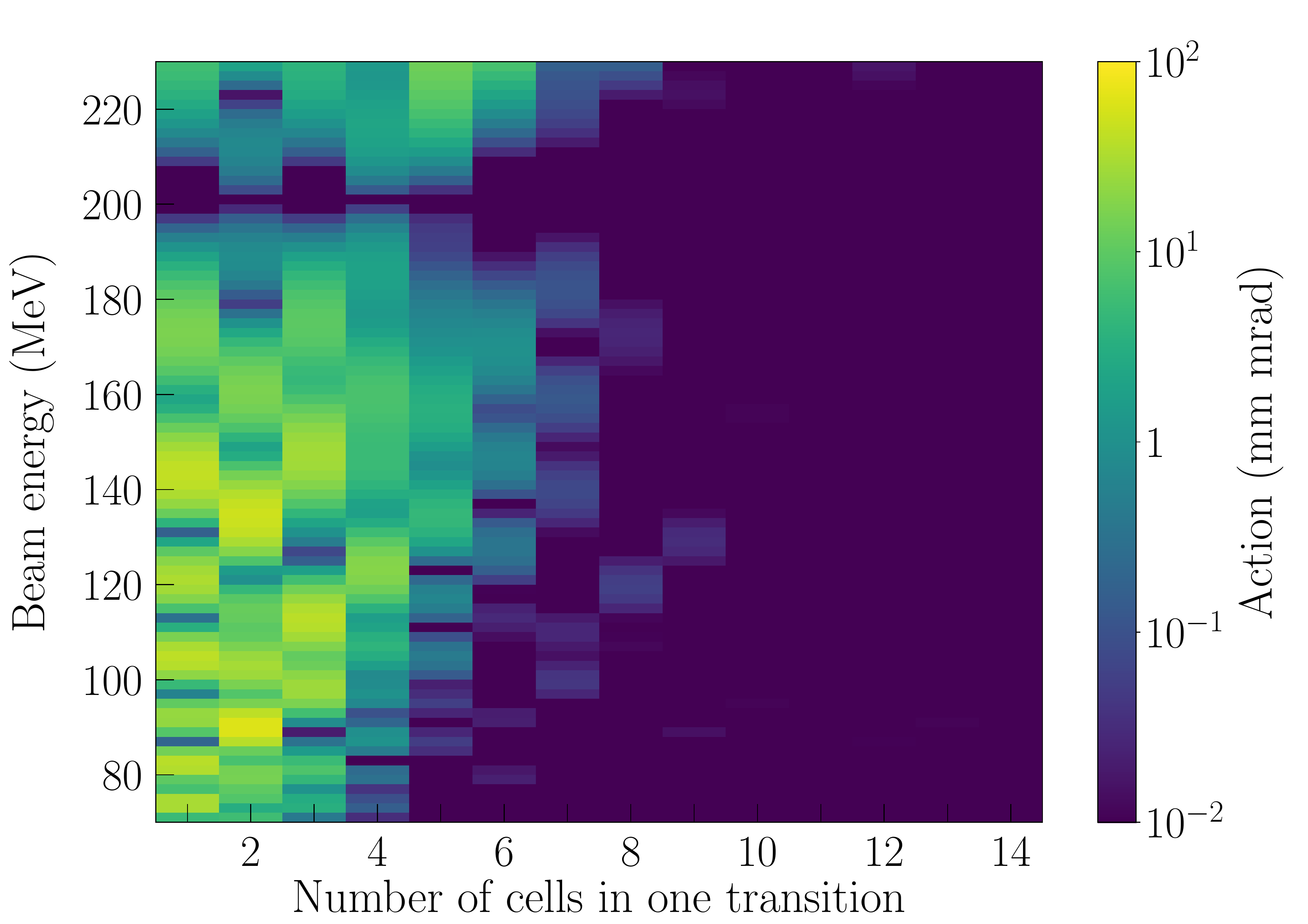}
    \caption{Normalised action in the x-plane at the end of the proposed gantry.}
    \label{fig:FG_action}
\end{figure}

\begin{figure}[t!]
    \centering
    \hspace*{-1em}
    \includegraphics[width=1.1\columnwidth]{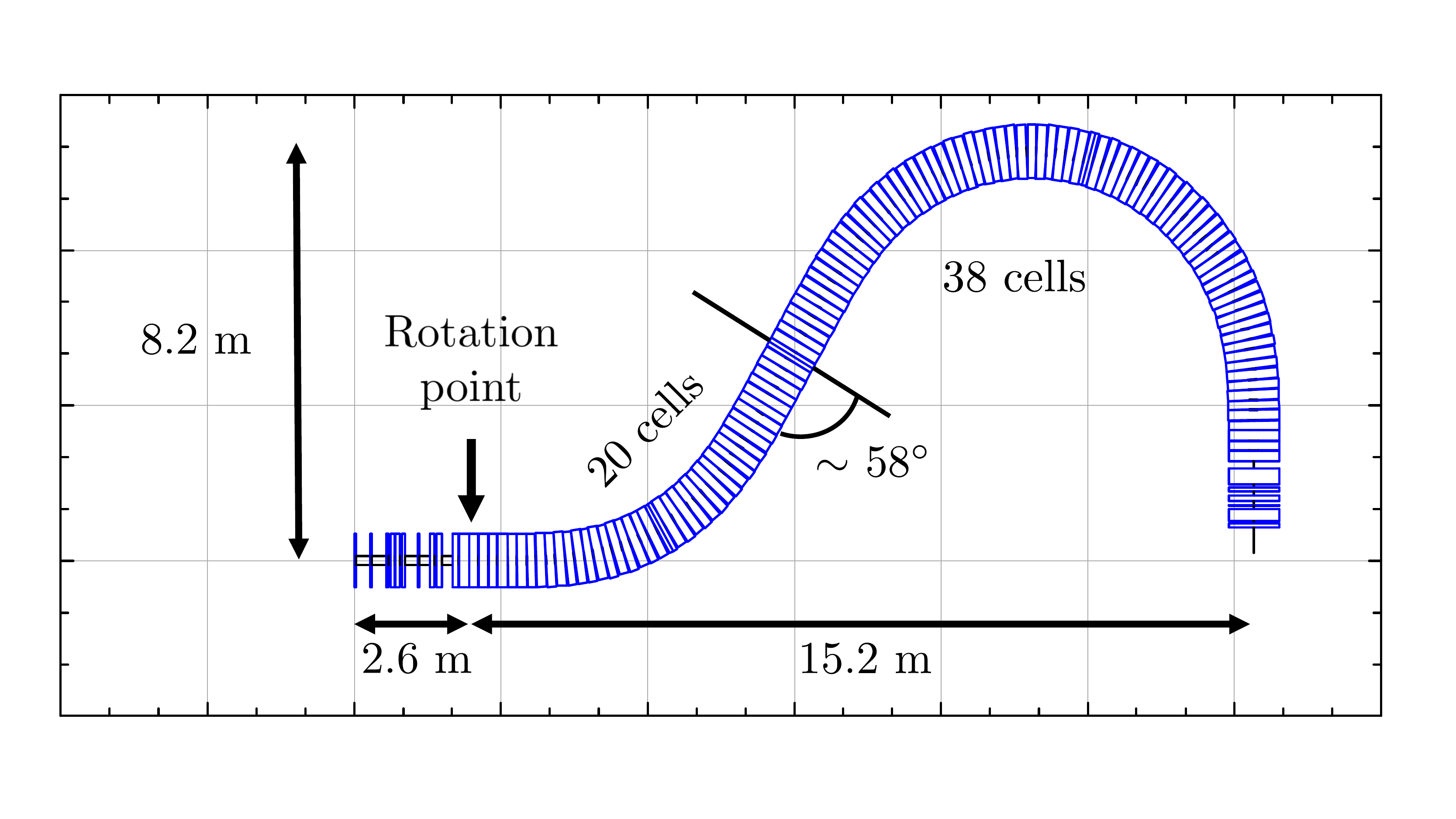}
    \vspace*{-2em}
    \caption{Dimensions and number of cells in the final design of the matching section and gantry.}
    \label{fig:FG_reprez}
\end{figure}

A separate analysis has been done by tracking a single particle from the axis of linac to the end of the gantry. The normalised action (\ref{eq:J_norm}) was calculated at the last cell of the gantry, treating the cell as periodic. Fig.~\ref{fig:FG_action} confirms that the transition sections limit the emittance growth if each section is composed of seven or more cells. The gantry is considered to be efficient if the action calculated at the last cell is smaller than the normalised emittance of the beam ($0.25$ mm mrad) for all energies. Based on the analysis of the transition sections and ensuring a suitable distance from the end cell to the isocentre, the final gantry layout is presented in Fig.~\ref{fig:FG_reprez}.

\vspace{0.3cm}
\textbf{Size:} Matching the entire energy range using an adiabatic transition results in an increase in the size of the gantry. The dimension obtained here are larger by a factor of almost 2 compared to previous designs which only match a discrete set of energies \cite{FFA17}. The rotating part requires a diameter of about 17 m and a length of 15 m compared to 11 m and 10 m for conventional, existing gantries \cite{Koschik:2015exq}. While the baseline is large, compromising the tight orbits or the small dispersion may bring down the overall size. Septum shaped magnets could be explored and the boundaries of field gradients may be pushed even higher.

\begin{figure*}[t!]
    \centering
    \hspace*{-1em}
    \includegraphics[width=1.1\textwidth]{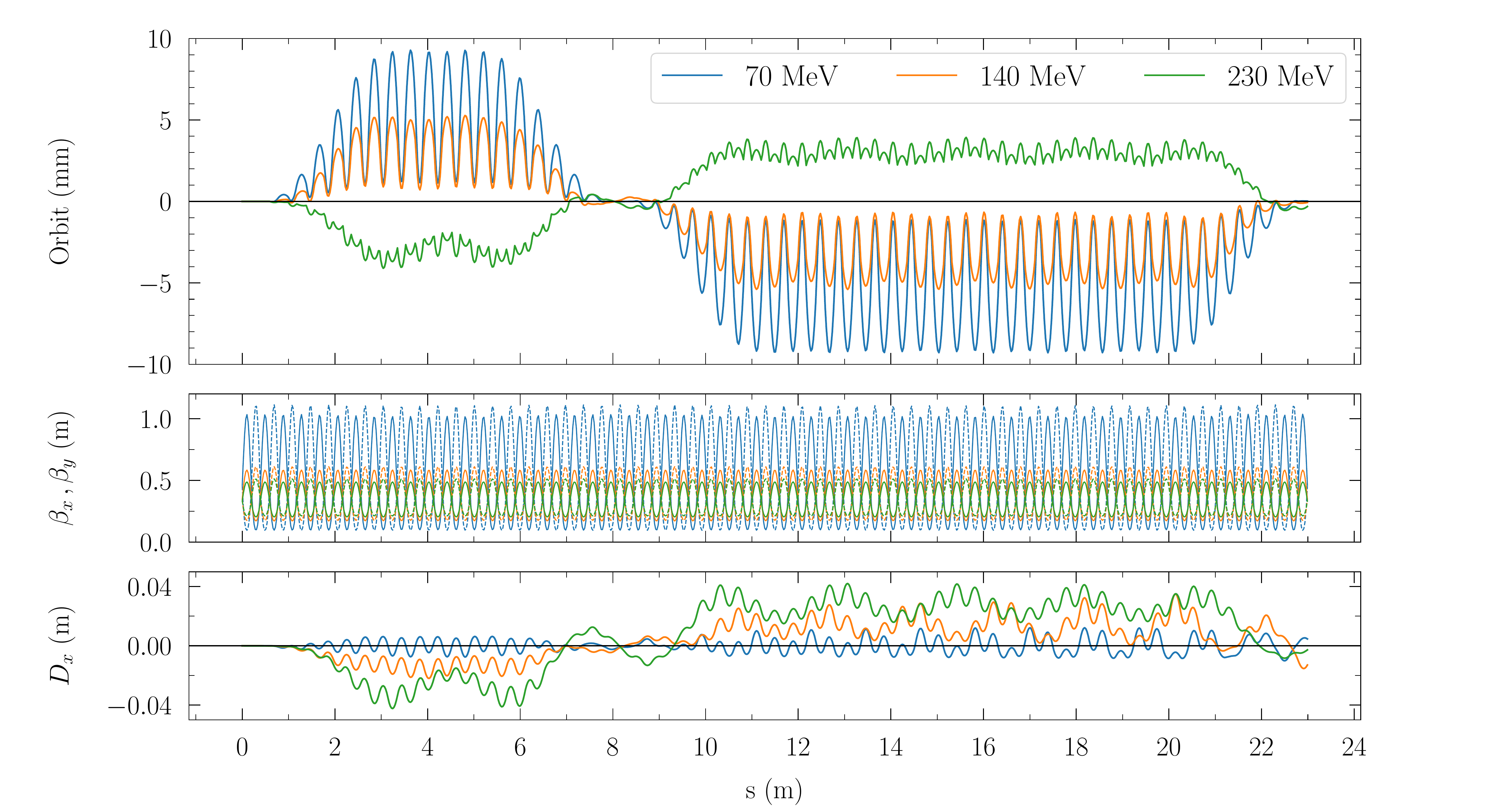}
    \caption{Orbits, betatron functions and dispersion for the final design of the gantry. Particles oscillate around the closed orbits and end up on the axis. Small dispersion is still remnant at the end.}
    \label{fig:FG_optics}
\end{figure*}

\vspace{0.3cm}
\textbf{Beam optics}: Fig.~\ref{fig:FG_optics} shows the particle orbits and the values of dispersion which extend over acceptable small ranges. There is still a non-zero dispersion at the end of the gantry which could not be decreased further. The gantry achieves good transport of the energy range onto the axis at the end and keeps the beam size small. The precision in the lateral beam position is less than 1 mm. The betatron functions remain close to the periodic solutions of each cell, keeping the beam size a factor of 10 smaller than previous results \cite{FFA18}.

\vspace{0.3cm}
\textbf{Weight:} The intrinsic lower weight of the magnets reduces the weight of the gantry by one or two orders of magnitude with respect to conventional gantries in operation. For the required field gradients and apertures of about 3\,cm diameter, the predicted outer diameter of the magnets is approximately\,8 cm (\ref{eq:Halbach}). The lighter components should also reduce the cost of the rotating structure.

\subsection{Tolerance to errors}

The reliability of such a gantry was investigated based on the tolerance against two types of machine errors: magnet alignment errors and errors in the field gradient of each magnet. For the bending elements, the error in the field gradient also introduces an error in the field strength. 

\begin{figure}[h!]
    \centering
    \hspace*{-1em}
    \includegraphics[width=0.51\textwidth]{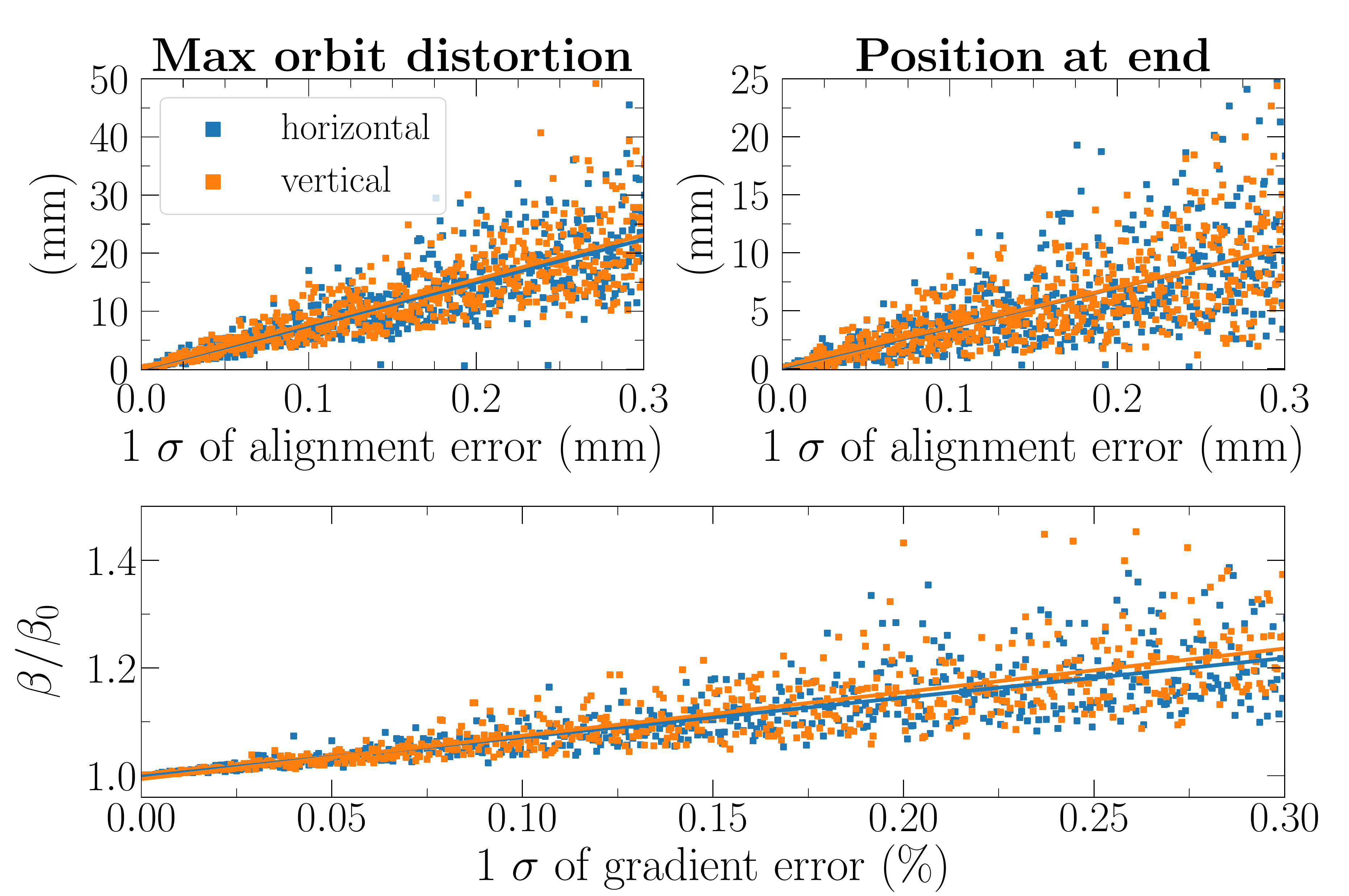}
    \caption{\textbf{Top}: Maximum orbit distortion and the largest particle position at the end of the gantry against alignment errors. \textbf{Bottom}: Maximum ratio of betatron function against field gradient errors. Straight lines indicate a linear fit.}
    \label{fig:errors}
\end{figure}

600 different iterations of the gantry were analysed for each type of error. For each design iteration, every magnet had an error sampled from a Gaussian distribution with $2 \, \sigma$ cut off \cite{PhysRevSTAB.13.084001}. The sigma value was varied from 0 to 0.3 mm for the alignment error and from 0\% to 0.3\% for the field gradient error. Fig.~\ref{fig:errors} shows that for practical purposes, the alignment errors have to be lower than 0.1\,mm, while the field quality can vary within 0.1\%. For each design iteration, the optics was calculated for 70\,MeV, 140\,MeV, and 230\,MeV from which the maximum deviation is presented.

\section{Conclusion}

We studied the use of several adiabatic transition sections in the design of a gantry optimised for delivering a proton beam from a linac. Matching the full energy range proved to be ineffective if the initial cell of the gantry is identical to the linac cell, due to their highly distinctive periodic orbit solutions. Alternatively, we found a suitable arc cell and optimised the transition towards a more compatible straight cell, adding a conventional matching section between the linac and the gantry. Overall limits on the size of the gantry were determined for matching the full energy range. This study may provide an initial assessment of this technology towards the planning and construction of linac based hadron therapy facilities.

Future research includes the addition of scanning magnets to the design and accounting for other factors in the optimisation, such as room or shielding size. Beam distortions need to be investigated to implement suitable correction schemes. Further improvements may be possible once the design is associated with the beam properties and the layout required by a particular user. With an industrial view, the automation of these design stages into one tool may be a valuable step forward by reducing the time invested in the planning of treatment facilities.

\section*{Acknowledgements}

We wish to thank Dr Alberto Degiovanni for the valuable discussions and gratefully acknowledge the insight and helpful suggestions from the ISIS Intense Beams Group at the Rutherford Appleton Laboratory.

\newpage

\addcontentsline{toc}{section}{References}
 \renewcommand{\refname}{References}
\bibliographystyle{unsrtnat-modif} 
\bibliography{main}{} 

\newpage
\section*{Appendix}
\subsection*{Further design specifications}

\begin{table}[H]
\caption{Parameters for the linac cell}
\label{table:linac_cell}
\vspace{-0.75cm}
\begin{center}
\begin{tabularx}{\columnwidth}{X r}
\hline
Total cell length (cm) & 66  \\ 
PMQs length (cm)  & 3 \\
Focusing gradient (T/m) & 102 \\
Defocusing gradient (T/m) & -102\\
Drift space length (cm) & 30 \\
Momentum acceptance $ \Delta p/p$ & $\pm$ 31\% \\
\hline
\end{tabularx}
\end{center}
\end{table}

\begin{table}[H]
\caption{Parameters for the FFA arc cell}
\label{table:arc_cell}
\vspace{-0.75cm}
\begin{center}
\begin{tabularx}{\columnwidth}{X r}
\hline
Momentum acceptance $ \Delta p/p$ & $\pm$ 31\% \\
Total cell length (mm) & 393  \\ 
Focusing quadrupole length (mm)  & 197 \\
Bending magnet length (mm) &  175 \\
Focusing gradient (T/m) & 92.41 \\
Defocusing gradient (T/m) & -102.98\\
Drift space length (mm) & 10 \\
Bend angle per cell (deg.) & 5 \\
Minimum orbit excursion (mm) & -9.2 \\
Maximum orbit excursion (mm) &  8.1 \\
\hline
\end{tabularx}
\end{center}
\end{table}

\end {document}